\documentclass[12pt]{article}
\usepackage{graphicx}
\usepackage{amsmath}
\usepackage{amssymb}
\usepackage{latexsym}

\def\Ket#1{\mathinner{|{#1}\rangle}}
\def\braket#1{\mathinner{\langle{#1}\rangle}}

\def\ket#1{\left|#1\right>}
{\catcode`\|=\active 
  \gdef\Braket#1{\left<\mathcode`\|"8000\let|\bravert {#1}\right>}}
\def\bravert{\egroup\,\vrule\,\bgroup}
\newcommand{\alg}[1]{#1}
\newcommand{\su}{\mathfrak{su}}
\newcommand{\so}{\mathfrak{so}}
\newcommand{\Sl}{\mathfrak{sl}}
\newcommand{\tr}{\mathop{\rm Tr}}
\newcommand{\bt}{{\tilde b}}
\newcommand{\at}{{\tilde a}}

\newcommand{\be}{\begin{eqnarray}}
\newcommand{\ee}{\end{eqnarray}}
\newcommand{\bea}{\begin{eqnarray}}
\newcommand{\eea}{\end{eqnarray}}
\newcommand{\ben}{\begin{equation}}
\newcommand{\een}{\end{equation}}

\newcommand{\nn}{\nonumber}
\newcommand{\atopfrac}[2]{\genfrac{}{}{0pt}{}{#1}{#2}}

\numberwithin{equation}{section}

\begin{document}

\begin{titlepage}
\begin{flushright}
CALT-68-2511\\
PUPT-2125\\
hep-th/0407096
\end{flushright}
\vspace{15 mm}
\begin{center}
{\huge  Lattice super Yang-Mills: \\
	A virial approach to operator dimensions }
\end{center}
\vspace{10 mm}

\begin{center}
{\large Curtis G.\ Callan, Jr.${}^{a}$,
Jonathan Heckman${}^{a}$,
Tristan McLoughlin${}^{b}$,\\
Ian Swanson${}^{b}$ }\\
\vspace{3mm}
${}^a$ Joseph Henry Laboratories\\
Princeton University\\
Princeton, New Jersey 08544, USA\\
\vspace{0.5 cm}
${}^b$ California Institute of Technology\\
Pasadena, CA 91125, USA 
\end{center}
\vspace{5 mm}
\begin{center}
{\large Abstract}
\end{center}
\noindent

The task of calculating operator dimensions in the planar limit of 
${\cal N}=4$ super Yang-Mills theory can be vastly simplified by 
mapping the dilatation generator to the Hamiltonian of an integrable spin chain.  
The Bethe ansatz has been used in this context to compute the spectra of spin 
chains associated with various sectors of the theory which are known to decouple 
in the planar (large-$N_c$) limit.  These techniques are powerful at leading order in 
perturbation theory but become increasingly complicated beyond one loop in the 
't~Hooft parameter $\lambda=g_{\rm YM}^2 N_c$, where spin chains typically acquire 
long-range (non-nearest-neighbor) interactions.  
In certain sectors of the theory, moreover, higher-loop Bethe ans\"atze
do not even exist. We develop a virial expansion of the spin chain Hamiltonian
as an alternative to the Bethe ansatz methodology, a method which simplifies the 
computation of dimensions of multi-impurity operators at higher loops in $\lambda$.
We use these methods to extract previously reported numerical gauge theory predictions
near the BMN limit for comparison with corresponding results on the string theory 
side of the AdS/CFT correspondence. For completeness, we compare our virial results
with predictions that can be derived from current Bethe ansatz technology.

\vspace{4mm}
\begin{flushleft}
\today
\end{flushleft}
\end{titlepage}
\newpage
\section{Introduction}

A two-dimensional parameter space emerges around the BMN limit of ${\cal N}=4$ 
$SU(N_c)$ super Yang-Mills (SYM) theory and the dual pp-wave
limit of IIB superstring theory on $AdS_5\times S^5$. 
This space is parameterized on one axis by the perturbative gauge theory expansion in 
powers of the 't~Hooft coupling $\lambda = g_{YM}^2 N_c$, and on the other 
by a string theory curvature expansion away from the Penrose limit in inverse 
powers of the string angular momentum $J$.  
The original comparison made by Berenstein, Maldacena and Nastase \cite{Berenstein:2002jq},
which has sparked a number of direct tests of the AdS/CFT correspondence in 
recent years, lies at the intersection of the one-loop order ($O(\lambda)$) 
gauge theory prediction and the zeroth-order ($O(J^{0})$) limit in the string 
curvature expansion.  To explore a larger region of this duality landscape,
a number of studies have pushed the string theory calculation to
higher orders in the $1/J$ curvature expansion 
\cite{Parnachev:2002kk,Callan:2003xr,Callan:2004uv}, 
while operator dimensions in the gauge theory have been computed 
to higher orders in $\lambda$ (see, eg.,
~\cite{Beisert:2003tq,Beisert:2003jb,Beisert:2003ys,Serban:2004jf,Beisert:2004hm}).

The first $1/J$ curvature correction to the fully quantized string theory
near the pp-wave limit of $AdS_5\times S^5$ was studied for
two-impurity string states in \cite{Parnachev:2002kk,Callan:2003xr,Callan:2004uv}.  
In this setting, the extended superconformal multiplet of the theory 
is a 256-dimensional multiplet built on a scalar primary 
(or ``highest-weight'') state.  This analysis was extended to the three-impurity, 
4,096-dimensional supermultiplet in \cite{Callan:2004ev}.  
To test predictions of the AdS/CFT correspondence, 
it was necessary in the course of the latter study to obtain higher-impurity, 
higher-loop operator dimensions from the gauge theory near the BMN limit. 
This is hard to do by standard diagrammatic methods, but the problem has been
drastically simplified by the recent discovery that, in certain sectors of the 
gauge theory, the dilatation operator can be mapped to the Hamiltonian of an 
integrable spin-chain system.  Calculating operator dimensions is therefore 
equivalent to finding the eigenvalue spectrum of these spin chains and certain
established techniques associated with integrable systems, most notably the
Bethe ansatz, have proved useful in this context (for a general review of the 
Bethe ansatz method, see \cite{Faddeev:1996iy}). The utility of this approach 
in the setting of ${\cal N}=4$ super Yang-Mills theory was first demonstrated by 
Minahan and Zarembo \cite{Minahan:2002ve}. For operators with two $R$-charge 
impurities, the spin chain spectra can be computed exactly via the Bethe ansatz.  
For three or higher-impurity operators, however, the Bethe equations have
(to the best of our knowledge) only been solved perturbatively near the limit 
of infinite chain length \cite{Minahan:2002ve,Beisert:2003yb,Lubcke:2004dg}. Furthermore, at 
higher-loop order in $\lambda$, the spin chain Hamiltonians typically acquire 
long-range or non-nearest-neighbor interactions for which a Bethe ansatz may not be 
available. For example, while the action of the spin chain Hamiltonian in the 
``closed $\su(2|3)$'' sector is known to three-loop order \cite{Beisert:2003ys}, 
the corresponding long-range Bethe ansatz is not known (though it may well exist).  
A long-range Bethe ansatz does exist for the particularly simple ``closed $\su(2)$'' 
sector of the theory \cite{Serban:2004jf,Beisert:2004hm}, and
our methods will provide a useful cross-check on these approaches
to higher-order gauge theory anomalous dimensions. 

To improve on the current limitations of Bethe ansatz techniques, we have developed 
a virial approach to the spin chain systems of ${\cal N}=4$ super Yang-Mills theory.
The generic spin-chain Hamiltonian acts on single-impurity pseudoparticles 
as a lattice Laplacian and higher $N$-body interactions among pseudoparticles
are suppressed  relative to the one-body pseudoparticle energy by inverse powers of 
the lattice length $L$.  Surprisingly, this expansion of the spin chain
Hamiltonian is truncated at $O(L^{-3})$ in certain subsectors of the theory, 
allowing straightforward eigenvalue calculations that are exact in the chain length for
operators with more than two $R$-charge impurities.  
Furthermore, since the goal is to eventually compare anomalous
dimensions with $1/J$ energy corrections to corresponding string states
near the pp-wave limit of $AdS_5\times S^5$, and because the string
angular momentum $J$ is proportional to the lattice length $L$, 
this virial expansion is precisely what is needed to devise a practical 
method for testing the AdS/CFT correspondence at any order in the gauge theory loop 
expansion for an arbitrary number of $R$-charge (or worldsheet) impurities.

The purpose of this paper is to provide a detailed 
description of these methods, and to compare previously
derived numerical predictions \cite{Callan:2004ev} with corresponding predictions from 
Bethe ansatz technology near the BMN limit. We will focus on three particular closed 
sectors of the theory, each labeled by the subalgebra of the full superconformal algebra 
which characterizes the spin variables of the equivalent spin chain system. 
Specifically, there are two sectors spanned by bosonic operators and
labeled by $\su(2)$ and $\Sl(2)$ subalgebras plus an $\su(2|3)$ sector which includes 
fermionic operators. Section 2 is dedicated to an analysis of the bosonic $\su(2)$ closed 
sector to three-loop order in $\lambda$. In section 3 we analyze an $\su(1|1)$  subsector 
of the closed $\su(2|3)$ system to three-loop order. The spin-chain Hamiltonian in the bosonic 
$\Sl(2)$ sector has previously been determined to one loop, and we analyze this system in 
section 4.  We conclude in the final section with a discussion of future 
applications and directions of study.

\section{The $\su(2)$ sector}

Single-trace operators in the closed $\su(2)$ sector are constructed from
two complex scalar fields of ${\cal N}=4$ SYM, typically 
denoted by $Z$ and $\phi$.  Under the 
$SO(6) \simeq U(1)_R\times SO(4)$ decomposition of the full
$SU(4)$ $R$-symmetry group, the $Z$ fields are charged under the
scalar $U(1)_R$ component and $\phi$ is a particular scalar field
carrying zero $R$-charge.  The basis of length-$L$ operators
in the planar limit is constructed from single-trace monomials with $I$ 
impurities and total $R$-charge equal to $L-I$: 
\be
\label{SO6basis}
\tr(\phi^I Z^{\alg L-I})\ , \qquad  
\tr(\phi^{I-1}Z\phi Z^{{\alg L-I}-1})\ , \qquad 
\tr(\phi^{I-2}Z\phi^2 Z^{{\alg L-I}-1})\ , \qquad \ldots
\ee
The statement that this sector of operators is ``closed'' means simply
that the anomalous dimension operator can be diagonalized on this
basis, at least to leading order in large $N_c$ \cite{Beisert:2003tq,Beisert:2003jj}.

The heart of the spin-chain approach is the proposition that there
exists a one-dimensional spin system whose Hamiltonian can be identified 
with the large-$N_c$ limit of the anomalous dimension operator acting
on this closed subspace of operators \cite{Minahan:2002ve}. Since the anomalous dimensions
are perturbative in the 't~Hooft coupling $\lambda$, it is natural 
to expand the $\su(2)$ spin chain Hamiltonian in powers of $\lambda$ as well:
\be
H_{\su(2)} = I + \sum_n \left( \frac{\lambda}{8 \pi^2} \right)^n H_{\su(2)}^{(2n)}\ .
\ee
Comparison with the gauge theory has shown that successive terms in the
expansion of the Hamiltonian have a remarkably simple structure:
the one-loop-order Hamiltonian $H^{(2)}_{\su(2)}$ is built out of 
permutations of pairs of nearest-neighbor fields and, at $n$-th order, the 
Hamiltonian permutes among themselves fields which are at most $n$ lattice 
sites apart. This is a universal structure which leads to remarkable 
simplifications in the various closed sectors of the theory \cite{Beisert:2003yb}.

Beisert, Kristjansen and Staudacher \cite{Beisert:2003tq} have introduced the
following useful notation for products of permutations acting on operators 
separated by an arbitrary number of lattice sites:
\be
\{ n_1,n_2,\dots\} = \sum_{k=1}^L P_{k+n_1,k+n_1+1} P_{k+n_2,k+n_2+1}\cdots\ ,
\ee
where $P_{i,j}$ simply exchanges fields on the $i^{th}$ and $j^{th}$ lattice sites 
on the chain. The upshot of the gauge theory analysis is that the 
equivalent spin-chain Hamiltonian for the $\su(2)$ sector can be written
in a rather compact form in terms of this notation. The result, correct to 
three-loop order, is (see \cite{Beisert:2003tq} for details)
\be
H_{\su(2)}^{(2)} & = & 2\left( \{ \}-\{0\} \right) 
\label{Hsu2_1}
\\
H_{\su(2)}^{(4)} & = & 2\bigl( -4\{\} + 6\{0\} - (\{0,1\} + \{1,0\}) \bigr)
\label{Hsu2_2}
\\
H_{\su(2)}^{(6)} & = & 4\bigl( 15\{\}-26\{0\} +6\left(\{0,1\}+\{1,0\}\right)
		+\{0,2\}-\left(\{0,1,2\}+\{2,1,0\}\right)
		\bigr)\ .
\label{Hsu2_3}
\ee
The form of the three-loop term $H_{\su(2)}^{(6)}$ was first conjectured in 
\cite{Beisert:2003tq} based on integrability restrictions and BMN scaling;
this conjecture was later corroborated by direct field-theoretic methods in 
\cite{Beisert:2003ys} (see also \cite{Beisert:2003jb} for relevant discussion
on this point). Our goal is to develop practical methods for finding the
eigenvalue spectrum of the spin-chain Hamiltonian for various interesting cases.

\subsection{One-loop order}
We start at one-loop order with $H_{\su(2)}^{(2)}$ in eqn.~(\ref{Hsu2_1}), which
provides a natural `position-space' prescription for constructing 
matrix elements in an $I$-impurity basis of operators.  As an explicit
example, we consider first the basis of two-impurity operators of length $L=8$:
\be
\tr(\phi^2 Z^6) \qquad 
\tr(\phi Z \phi Z^5)  \qquad 
\tr(\phi Z^2 \phi Z^4)  \qquad 
\tr(\phi Z^3 \phi Z^3)\ .
\ee 
It is easy to see that the one-loop Hamiltonian mixes the four elements of 
this basis according to the matrix
\be
H_{\su(2)}^{(2)} = \left(
\begin{array}{cccc}
2 & -2 & 0 & 0 \\
-2 & 4 & -2 & 0 \\
0 & -2 & 4 & -2\sqrt{2} \\
0 & 0 & -2\sqrt{2} & 4 
\end{array}\right)\ .
\label{L6PS}
\ee
This tri-diagonal matrix generalizes to arbitrary $L$ and it is simple to show that
the two-impurity one-loop eigenvalues of $H_{\su(2)}^{(2)}$ are given by the formula 
\cite{Beisert:2002tn}
\be
E_{\su(2)}^{(2)} = {8}\, \sin^2 \left(\frac{\pi n}{L-1}\right) \qquad
n = 0,\ldots,n_{\rm max} = 
	\biggl\{ \genfrac{}{}{0pt}{0}{ (L-2)/2, \quad  L\ {\rm even} }{ (L-3)/2, \quad  L\ {\rm odd} }\ .
\label{su2EIG}
\ee
Although we defer our discussion of the Bethe ansatz approach until later
in this section, we note that the two-impurity $\su(2)$ Bethe equations for this 
spin chain \cite{Minahan:2002ve} can be solved exactly and their solution
agrees with eqn.~(\ref{su2EIG}).  Note that if the denominator $L-1$ were replaced by 
$L$, the above expression would agree with the usual lattice Laplacian energy for a 
lattice of length $L$. The difference amounts to corrections to the free Laplacian of 
higher order in $1/L$ and we will seek to understand the physical origin of such 
corrections in what follows.

To compare gauge theory predictions with $1/J$ corrections to the 
three-impurity spectrum of the string theory on $AdS_5\times S^5$,
we need to determine the large-$L$ behavior of the three-impurity
spin chain spectrum.  We are primarily interested in systems with few 
impurities compared to the length of the spin chain and we expect that
impurity interaction terms in the Hamiltonian will be suppressed by
powers of the impurity density (i.e.~inverse powers of the lattice length).
This suggests that we develop a virial expansion of the spin 
chain Hamiltonian in which the leading-order term in $1/L$ gives the 
energy of free pseudoparticle states on the lattice (labeled by lattice
momentum mode numbers as in the two-impurity spectrum eqn.~(\ref{su2EIG})) and higher $1/L$
corrections come from $N$-body interactions described by vertices $V_N$. 
A reasonable guess about how the $N$-body interactions should scale with $1/L$
suggests that we can write the one-loop-order energy for $I$ impurities
in the form 
\be
E(\{n_i\}) = I + \frac{\lambda}{2\pi^2}\sum_{i=1}^I \sin^2\frac{n_i\pi}{L}
+ \sum_{N=2}^{2I} \frac{\lambda}{L^{2N-1}} {V_{N-\rm body}(n_1,\ldots,n_I)} + \cdots\ ,
\label{Nbody}
\ee
where the leading-order contribution $I$ measures the naive dimension
minus $R$-charge, the next term is the lattice Laplacian energy of $I$
non-interacting pseudoparticles and the $1/L$ corrections account for 
interactions between pseudoparticles (which may depend on the lattice 
momenta mode numbers $n_i$). In the many-body approach, one would try
to derive such energy expressions by rewriting the Hamiltonian in terms of 
creation/annihilation operators $b_{n_i}$, $b_{n_i}^\dag$ for the
pseudoparticles (commuting or anticommuting as appropriate). The $N$-body 
interaction vertex would generically be written in terms of the $b,b^\dag$ as
\be
V_N = \sum_{n_i,m_i}~\delta_{n_1+\cdots+n_N,m_1+\cdots+m_N}
	f_N(\{n_i\},\{m_i\})\prod_{i=1}^{N} b_{n_i}^\dag \prod_{i=1}^{N} b_{m_i}\ ,
\ee
where $f_N(\{n_i\},\{m_i\})$ is some function of the lattice momenta 
and the Kronecker delta enforces lattice momentum conservation. One has to
determine the functions $f_N$ by matching the many-body form of the Hamiltonian
to exact spin chain expressions such as eqn.~(\ref{Hsu2_1}). We will see that,
once the Hamiltonian is in many-body form, it is straightforward to obtain
a density expansion of the higher-impurity energy eigenvalues.

The discussion so far has been in the context of one-loop gauge theory
physics, but the logic of the virial expansion should be applicable
to the general case. To include higher-loop order physics we must do
two things: a) generalize the functions $f_N(\{n_i\},\{m_i\})$ defining 
the multi-particle interaction vertices to power series in $\lambda$ and
b) allow the free pseudoparticle kinetic energies themselves to become
power series in $\lambda$. We will be able to carry out the detailed construction 
of the higher-loop virial Hamiltonian in a few well-chosen cases.
To match this expansion at $n$-loop order in $\lambda$ to the corresponding
loop order (in the modified 't~Hooft coupling $\lambda' = g_{\rm YM}^2 N_c / J^2$) 
in the string theory, we need to determine the Hamiltonian to $O(L^{-(2n+1)})$ 
in this virial expansion.  (The first curvature correction to the pp-wave string theory 
at one loop, for example, appears at $O(\lambda'/J)$ or, in terms of gauge theory 
parameters, at $O(\lambda/L^3)$.) Auspiciously, it will turn out that this virial 
expansion in the $\su(2)$ sector is truncated at small orders in $1/L$, allowing 
for simple eigenvalue calculations that are exact in $L$ (although perturbative
in $\lambda$).

The first step toward obtaining the desired virial expansion is to recast the spin 
chain Hamiltonian $H_{\su(2)}$, which is initially expressed in terms of permutation 
operators, in terms of a creation and annihilation operator algebra. 
We begin by introducing the spin operators
\be
S^\pm = \frac{1}{2}\left( \sigma_x \pm i \sigma_y \right) \qquad
S^z = \frac{1}{2} \sigma_z\ ,
\ee
where $\vec \sigma$ are the Pauli matrices and $S^\pm_j,\ S^z_j$ act on 
a two-dimensional spinor space at the $j^{th}$ lattice site in the chain.
In this setting the $Z$ and $\phi$ fields are understood to be modeled
by up and down spins on the lattice. The nearest-neighbor permutation operator 
$P_{i,i+1}$ can be written in terms of spin operators as
\be
P_{i,i+1} = S_i^+S_{i+1}^- + S_i^-S_{i+1}^+ + 2S_i^zS_{i+1}^z + \frac{1}{2}\ , 
\ee
and the one-loop Hamiltonian in eqn.~(\ref{Hsu2_1}) can be written as
\be
H_{\su(2)}^{(2)} = - \sum_{j=1}^L 
	\left( S^+_j S^-_{j+1} + S^-_j S^+_{j+1}\right)
	-2 \sum_{j=1}^L S_j^z S_{j+1}^z + \frac{1}{2}\ .
\label{HS1}
\ee
A Jordan-Wigner transformation can now be used to express the spin generators
in terms of anti-commuting creation and annihilation operators (anti-commuting
because each site can be either unoccupied ($Z$) or occupied once ($\phi$)).
A pedagogical introduction to this technique can be found in \cite{Nagaosa:1999uc}.
The explicit transformation is
\be
\label{JordWig}
S_j^+ & = & b_j^\dag K(j) = K(j) b_j^\dag \nn\\
S_j^- & = & K(j) b_j = b_j K(j)  \nn\\
S_j^z & = & b_j^\dag b_j - {1}/{2}\ ,
\ee
where the Klein factors
\be
K(j) = \exp \left( i\pi \sum_{k=1}^{j-1} b_k^\dag b_k \right)
\ee
serve to ensure that spin operators on different
sites commute, despite the anticommuting nature of the $b_j$.
The functions $K(j)$ are real, Abelian and, for $j\leq k$, 
\be
[ K(j),{\bf S}_k ] = 0\ . 
\ee
The operators $b_j$ and $b_j^\dag$ can therefore be written as
\be
b_j^\dag = S_j^+ K(j) \qquad b_j = S_j^- K(j)\ ,
\ee
and we easily verify that they satisfy the standard anticommutation relations
\be
\{ b_j, b_k^\dag \} = \delta_{jk} \qquad \{b_j^\dag,b_k^\dag\} = \{b_j,b_k\} = 0\ .
\ee
Cyclicity on the lattice requires that ${\bf S}_{L+1} = {\bf S}_1$, a condition
which can be enforced by the following  boundary condition on the creation and annihilation operators:
\be
b_{L+1} = (-1)^{I+1} b_1 \qquad I \equiv \sum_{j=1}^L b_j^\dag b_j\ ,
\label{BC}
\ee
where the integer $I$ counts the number of spin chain impurities.
Since we are primarily interested in the three-impurity problem 
(for comparison with the corresponding string results reported in \cite{Callan:2004ev}),
we will henceforth impose the boundary conditions in eqn.~(\ref{BC}) for 
odd impurity number only. We can use all of this to re-express eqn.~(\ref{HS1}) 
in creation and annihilation operator language, with the result
\be
H_{\su(2)}^{(2)} =   \sum_{j=1}^{L} \left(
	b_{j}^\dag b_{j} + b_{j+1}^\dag b_{j+1} - b_{j+1}^\dag b_{j} - b_{j}^\dag b_{j+1}
	+ 2\,b_{j}^\dag b_{j+1}^\dag b_{j} b_{j+1}  \right)\ .
\label{Hsu2PS}
\ee
Converting to momentum space via the usual Fourier transform
\be
b_j = \frac{1}{\sqrt{L}}\sum_{p=0}^{L-1} e^{ -\frac{2\pi i j}{L} p }~\bt_p\
\ee
yields
\be
H_{\su(2)}^{(2)} & = & 4\sum_{p=0}^{L-1} \sin^2\left( \frac{\pi p}{L} \right) \bt_p^\dag \bt_p
	+ \frac{2}{L} \sum_{p,q,r,s=0}^{L-1} e^{\frac{2\pi i (q-s)}{L}} 
	\bt_p^\dag \bt_q^\dag \bt_r \bt_s\, \delta_{p+q,r+s}\ .
\label{Hsu2MS}
\ee
This is a rather standard many-body Hamiltonian: it acts on a Fock space
of momentum eigenstate pseudoparticles, contains a one-body pseudoparticle 
kinetic energy term and a two-body pseudoparticle interaction (the latter
having the critical property that it conserves the number of pseudoparticles).
Note that the Hamiltonian terminates at two-body interactions, a fact which will
simplify the virial expansion of the energy spectrum. This termination is a 
consequence of the fact that the one-loop Hamiltonian contains only 
nearest-neighbor interactions and that lattice sites can only be once occupied.

Because the pseudoparticle (or impurity) number is conserved by the interaction,
three-impurity eigenstates of the Hamiltonian must lie in the space spanned by
\be
\bt_{k_1}^\dag \bt_{k_2}^\dag \bt_{k_3}^\dag \ket{L}  \qquad
	k_1 + k_2 + k_3 = 0 \mod L\ ,
\ee
where the ground state $\ket{L}$ is identified with the zero-impurity operator $\tr(Z^L)$
and the condition of vanishing net lattice momentum arises from 
translation invariance on the spin chain (which in turn arises from the cyclicity
of the  single-trace operators in the operator basis). As a concrete example, 
the basis of three-impurity states of the $L=6$ $\su(2)$ spin chain is
\be
\bt_0^\dag \bt_1^\dag \bt_5^\dag \ket{L} \qquad 
\bt_0^\dag \bt_2^\dag \bt_4^\dag \ket{L} \qquad 
\bt_1^\dag \bt_2^\dag \bt_3^\dag \ket{L} \qquad 
\bt_3^\dag \bt_4^\dag \bt_5^\dag \ket{L}\ ,
\ee
and the matrix elements of the Hamiltonian (\ref{Hsu2MS}) 
in this basis are easily computed:
\be
H_{\su(2)}^{(2)} = \left(
\begin{array}{cccc}
\frac{1}{3} & -1 &  \frac{1}{3}& \frac{1}{3} \\
-1 & 3 & -1 & -1 \\
\frac{1}{3} & -1 &\frac{19}{3}  & \frac{1}{3} \\
\frac{1}{3} & -1 & \frac{1}{3} &  \frac{19}{3}
\end{array}\right)\ .
\label{L6MS}
\ee
The first-order perturbation theory corrections to the 
three-impurity operator anomalous dimensions are the eigenvalues of
this matrix.

\begin{figure}[htb]
\begin{center}
\includegraphics[width=3.5596in,height=2.2in,angle=0]{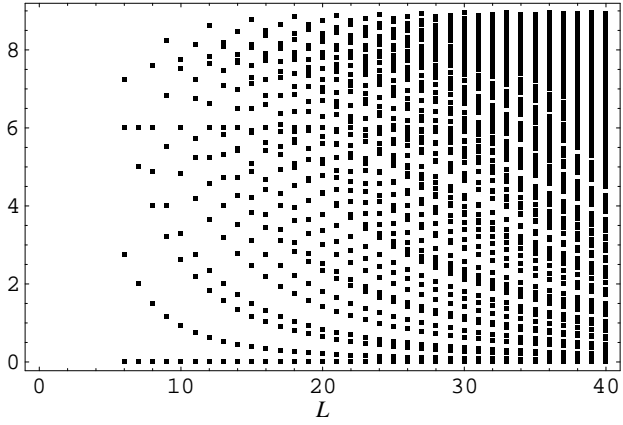}
\caption{One-loop $\su(2)$ spin chain spectrum vs.~lattice length $L$ ($6\leq L \leq 40$)}
\label{SO63aSTR}
\end{center}
\end{figure}

The construction and diagonalization of the Hamiltonian matrix on
the degenerate basis of three-impurity operators can easily be carried
out for larger $L$. The results of doing this\footnote{Using the position- 
or momentum-space formalism is purely a matter of 
convenience.  In practice we have found that for all sectors the momentum-space 
treatment is computationally much more efficient. The large-$L$ extrapolations
of both methods can be checked against each other, and we of course find
that they are in agreement.}
for lattice sizes out to $L=40$ are displayed in figure~\ref{SO63aSTR}.  
According to eqn.~(\ref{Nbody}), we expect the eigenvalues of $H_{\su(2)}^{(2)}$ 
to scale for large $L$ according to
\be
\label{scaling}
E_{\alg L}(\lbrace k_i \rbrace) =  \frac{\lambda}{\alg L^2}{E^{(1,2)}
	(\lbrace k_i \rbrace)} + 
\frac{\lambda}{\alg L^3}{E^{(1,3)}(\lbrace k_i \rbrace)} + O(\lambda \alg L^{-4})\ .
\label{expG1}
\ee
The scaling coefficients $E^{(1,2)}_{\su(2)}$ and $E^{(1,3)}_{\su(2)}$ can easily be extracted from 
the data displayed in figure~\ref{SO63aSTR} by fitting the spectral curves to 
large-order polynomials in $1/L$ (a similar treatment was used in \cite{Beisert:2003ea}). 
The results of this procedure are recorded for several 
low-lying levels in the spectrum (excluding zero eigenvalues) in table~\ref{NUM_SU(2)_1Loop}. 
\begin{table}[ht!]
\begin{eqnarray}
\begin{array}{|cccc|}
\hline
E_{\su(2)}^{(1,2)} & E_{\su(2)}^{(1,3)} & E_{\su(2)}^{(1,3)}/E_{\su(2)}^{(1,2)}  
                                  & {\rm Lattice\ Momenta\ }(k_1,k_2,k_3) \\
\hline
1+2.6\times 10^{-9}	&2-4.9\times 10^{-7}	&2-5.0\times 10^{-7}	& (1,0,-1)	 \\
3+4.6\times 10^{-9}	&7-8.8\times 10^{-7}    &7/3-3.0\times 10^{-7} & (1,1,-2)	 \\	
3+4.6\times 10^{-9}	&7-8.8\times 10^{-7}	&7/3-3.0\times 10^{-7}	&  (-1,-1,2)	 \\
4+6.0\times 10^{-9}	&8-1.1\times 10^{-6}    &2-2.9\times 10^{-7}   &  (2,0,-2)	 \\
7+3.2\times 10^{-8}	&14-7.1\times 10^{-6}	&2-1.0\times 10^{-6}	&  (1,2,-3) 	 \\
7+3.2\times 10^{-8}	&14-7.1\times 10^{-6}	&2-1.0\times 10^{-6}	&  (-1,-2,3)	 \\
9+2.2\times 10^{-7}	&18-5.1\times 10^{-5}	&2-5.7\times 10^{-6}	&  (3,0,-3)	 \\
12+5.7\times 10^{-5}	&28+3.8\times 10^{-3}	&7/3-1.4\times 10^{-3}	&  (2,2,-4)	 \\
12+5.7\times 10^{-5} 	&28+3.8\times 10^{-3}	&7/3-1.4\times 10^{-3}	&  (-2,-2,4)	 \\
13-5.6\times 10^{-5}	&26-3.8\times 10^{-3}	&2+1.3\times 10^{-3}	&  (1,3,-4) 	 \\
13-5.6\times 10^{-5}	&26-3.8\times 10^{-3}	&2+1.3\times 10^{-3}	&  (-1,-3,4)	 \\
\hline
\end{array} \nonumber
\end{eqnarray}
\caption{Scaling limit of three-impurity
	$\su(2)$ numerical spectrum at one loop in $\lambda$}
\label{NUM_SU(2)_1Loop}
\end{table}
As originally reported in \cite{Callan:2004ev}, string theory makes  the
following simple predictions for the large-$L$ $\su(2)$ expansion coefficients 
$E_{\su(2)}^{(1,3)}$ and $E_{\su(2)}^{(1,2)}$: 
\be
E_{\su(2)}^{(1,2)} = (k_1^2+k_2^2+k_3^2)/2 & \qquad & k_1+k_2+k_3=0\nonumber\nonumber\\
E_{\su(2)}^{(1,3)}/E_{\su(2)}^{(1,2)}  =  2 & \qquad & (k_1\neq k_2\neq k_3)
\nonumber\\
E_{\su(2)}^{(1,3)}/E_{\su(2)}^{(1,2)} = \frac{7}{3} & \qquad &(k_1 = k_2,\ k_3=-2k_1)\ .
\label{su2num}
\ee
Note the slight annoyance that we must distinguish the case where all mode indices are 
unequal from the case where two indices are equal and different from the third.
The last column of table~\ref{NUM_SU(2)_1Loop} displays the choice of indices
$\{k_i\}$ that best fit each spectral series and the other columns 
display the deviation of the extrapolation coefficients from the string
theory predictions of eqn.~(\ref{su2num}).  As the lattice momenta
increase, higher-order $1/L$ corrections to the spectrum become stronger
and more data will be required to maintain a given level of 
precision of the polynomial fit.  
This effect can be seen directly in the extrapolated eigenvalues in 
table~\ref{NUM_SU(2)_1Loop}. Nonetheless, it is clear from the table that the gauge 
theory match to the string theory prediction is extremely good.

We also note that the spectrum in table~\ref{NUM_SU(2)_1Loop} exhibits 
a degeneracy of eigenstates whose momentum labels are related by an overall
sign flip (a symmetry that is implemented on the operator basis by a parity 
operator $P$ which reverses the ordering of all fields within the trace). 
This degeneracy among ``parity pairs'' of gauge theory operators
was observed in \cite{Beisert:2003tq}, where it 
was shown that it arises as a consequence of integrability (which can, in turn, 
be used to constrain the form of the Hamiltonian at higher loop order 
\cite{Beisert:2003jb}). See \cite{Swanson:2004mk} 
for further discussion on the implications of this degeneracy.

To corroborate these results on the gauge theory side we turn to the one-loop 
Bethe ansatz for the Heisenberg spin chain. The Bethe ansatz for chains of 
spins in arbitrary representations of arbitrary simple Lie groups was developed
some time ago \cite{Ogievetsky:hu} (see also \cite{Saleur:1999cx} for an extension to
supersymmetric spin chains) and applied only recently to the specific
case of the dilatation operator of ${\cal N}=4$ SYM \cite{Minahan:2002ve,Beisert:2003yb}.
In the notation of \cite{Beisert:2003yb}, the Bethe equations are expressed in 
terms of the so-called Bethe roots 
(or rapidities) $u_{i}$ associated with the various impurity insertions 
in the single-trace ground state $\tr Z^L$.  In a one-dimensional dynamical 
interpretation, the impurities are pseudoparticle excitations and the roots 
parameterize in some fashion the lattice momenta of the pseudoparticles. 
The index $i$ in the Bethe root $u_i$ runs over the total number $I$ of 
impurities.  A second index $q_i = 1,\ldots ,7$ is used to associate each of the
$I$ Bethe roots with a particular simple root of the $\Sl(4|4)$ symmetry
algebra associated with ${\cal N}=4$ SYM.   
The Bethe ansatz then takes the form (see \cite{Beisert:2003yb} and references therein 
for further details)
\be
\left( \frac{u_{i} + \frac{i}{2}V_{q_i}}{u_{i}-\frac{i}{2}V_{q_i} }\right)^L
	= \prod_{j\neq i}^{I}
	\left(
	\frac{u_{i}-u_{j}+\frac{i}{2}M_{q_i,q_j}}
	     {u_{i}-u_{j}-\frac{i}{2}M_{q_i,q_j}}
	\right)\ ,
\label{BAEFull}
\ee
where $V_{q_i}$ denotes the ${q_i}^{th}$ Dynkin coefficient of the 
spin representation and $M$ is the Cartan matrix of the algebra.
To be slightly more specific, if $\alpha_{q_i}$ are the root vectors
associated with the nodes of the Dynkin diagram and $\mu$ is the
highest weight of the spin representation, then the Dynkin coefficient (for a bosonic algebra) 
is $V_{q_i} = 2\,\alpha^{(q_i)}\cdot \mu / (\alpha^{(q_i)})^2$ and the elements of
the Cartan matrix are $M_{q_i,q_j} = 2\,\alpha^{(q_i)}\cdot \alpha^{(q_j)} / (\alpha^{(q_j)})^2$
(note that diagonal elements $M_{q_i,q_i}=2$). 
(For superalgebras see, eg.,~\cite{LSA1,LSA2}.)
Furthermore, since the spin chain systems of interest to us are cyclic and 
carry no net momentum (analogous to the level-matching condition in the string 
theory), the Bethe roots $u_{i}$ are subject to the additional constraint
\be
1 = \prod_{i}^{I}\left(
	\frac{u_{i} + \frac{i}{2}V_{q_i}}{u_{i}-\frac{i}{2}V_{q_i}}
	\right)\ .
\label{BAE2}
\ee
Finally, having found a set of Bethe roots $u_{i}$ that solve the above
equations, the corresponding energy eigenvalue 
(up to an overall additive constant; see, eg.,~\cite{Beisert:2003yb}) is given by
\be
E 
= \sum_{j=1}^{I} \left( \frac{V_{q_j}}{u_j^2+V_{q_j}^2/4}\right)\ .
\label{betheenergy1L}
\ee
In the current application all impurities are of the same type 
(i.e.~carry the same Dynkin label), so the index $q_i$ can be ignored. 
It is worth noting, however, that the Dynkin coefficient $V_{q_i}$ can vanish, in which
case the associated Bethe roots do not contribute directly to the energy.

The Bethe equations are typically exactly soluble for the case of
two identical impurities (i.e.~two Bethe roots $u_{1}$, $u_{2}$ associated
with the same simple root of the algebra). The two-impurity $\su(2)$ Bethe equations, 
for example, yield solutions
that reproduce the familiar two-impurity anomalous dimension formula
noted above in eqn.~(\ref{su2EIG}) (see \cite{Minahan:2002ve,Beisert:2003yb} 
for further examples).   For three and higher impurities, however, 
exact solutions are not known. Since we are 
ultimately interested in comparing with string theory predictions at large 
values of the $S^5$ angular momentum $J$, an alternate approach is to solve 
the Bethe equations perturbatively in small $1/L$. Experience
shows that, in the limit where we can neglect interactions between
excitations (or impurities), the Bethe roots are simply the inverse
of the conserved momentum carried by the impurities. 
With a little work, one can show that the Bethe ansatz conditions,
eqns.~(\ref{BAEFull},\ref{BAE2}), can be solved order-by-order in a 
large-$L$ expansion:
\be
u_{i} = \frac{1}{2\pi k_{i}}\left(L + A_{i}\sqrt{L} + B_{i} + \cdots\right)\ ,
\label{uexp}
\ee
where $0<k_i<L$ is the usual integer lattice momentum. The half-integer powers 
of $L$ may or may not be present in eqn.~(\ref{uexp}): they are needed to deal with special kinematic 
situations (such as when a pair of impurities have the same lattice momentum) where
the integral power expansion would be singular.
The eigenvalues of the spin chain 
(or the anomalous dimensions of the corresponding gauge theory operator) 
are then obtained as a power series in $1/L$ by substituting the expansion
of the Bethe roots into eqn.~(\ref{betheenergy1L}).
This is the approach introduced by Minahan and Zarembo
for the $\so(6)$ spin chain in \cite{Minahan:2002ve}.
Since we wish to carry out similar calculations at higher orders in $\lambda$, 
we will review this methodology at one-loop order for the specific case of three 
identical impurities in the $\su(2)$ spin chain.  
(Since the $\su(2)$ chain is a subsector of the $\so(6)$ system studied in
\cite{Minahan:2002ve}, the three-impurity Bethe ansatz predictions derived here
are of course implied by the all-impurity $\so(6)$ anomalous dimension formula 
derived in \cite{Minahan:2002ve} at one loop.)

We now apply this to the closed $\su(2)$ sector where the Dynkin diagram has
a single node, the Cartan matrix is $M_{\su(2)} = 2$ and the Dynkin coefficient
of the fundamental representation is $V_{\su(2)} = 1$. Consequently, the Bethe equations
(\ref{BAEFull},\ref{BAE2}) reduce to
\be
&&\kern-25pt
	\left(\frac{u_{i}+i/2}{u_{i}-i/2}\right)^L
	= \prod_{j\neq i}^{I}\left(
	\frac{u_{i} - u_{j} + i}{u_{i}-u_{j} - i}
	\right)
\label{BAESU2_1}
\\
&&
\kern+10pt	1 = \prod_{i}^{I}\left(
	\frac{u_{i} + {i}/{2}}{u_{i}- {i}/{2}}
	\right)\ .
\label{BAESU2_2}
\ee
With three or more pseudoparticle excitations, bound-state solutions
can arise which satisfy the second equation (\ref{BAESU2_2}).
These solutions are characterized as having
pseudoparticle states sharing the same lattice momenta 
(eg.~$k_i = k_j$ for the $i^{th}$ and $j^{th}$ roots).
The generic solutions to the Bethe equations can therefore be
loosely divided into those which do or do not contain bound 
states.  For three impurities with no bound states present
($k_1 \neq k_2 \neq k_3$), eqn.~(\ref{BAESU2_2}) states that
$k_3 = -k_1 -k_2$. The strategy of \cite{Minahan:2002ve} can then be
used to obtain a systematic expansion of $\su(2)$ Bethe roots in powers
of $L^{-1}$, with the result
\be
u_1 & = & \frac{L-4}{2\pi k_1} + \frac{3k_1}{\pi(k_1-k_2)(2k_1+k_2)} + O(L^{-1})
\nn\\
u_2 & = & \frac{(L-4)k_1^2+(L-4)k_1k_2-2(L-1)k_2^2}{2\pi k_2(k_1^2+k_1 k_2 -2k_2^2)} + O(L^{-1})
\nn\\
u_3 & = & -\frac{(L-1)k_1^2-(8-5L)k_1k_2+2(L-1)k_2^2}{2\pi (k_1+k_2)(2k_1+k_2)(k_1+2k_2)} + O(L^{-1})  \ .
\ee
Substituting these roots into the energy formula eqn.~(\ref{BAE2}) gives the following
expression for the anomalous dimension of the $\su(2)$ three-impurity operator
at one-loop:
\be
E_{\su(2)}^{(2)}(k_1,k_2) = \frac{8 \pi^2}{L^3}\left(k_1^2 + k_1 k_2 + k_2^2\right)\left(L+2\right)
	+ O(L^{-4})
\qquad (k_1 \neq k_2 \neq k_3)\ .
\label{BAsu2}
\ee
This is in perfect agreement with the string theory results of eqn.~(\ref{su2num}) and, of 
course, the numerical gauge theory results in table~\ref{NUM_SU(2)_1Loop}.
When a single bound state is present the Bethe roots must be altered.  Taking,
for example, $k_1 = k_2$, the cyclic constraint in eqn.~(\ref{BAESU2_2})
sets $k_3 = -2k_1$, and the Bethe roots are
\be
u_1  & = & \frac{-7+3i\sqrt{L}+3L}{6\pi k_1}+ O(L^{-1/2}) 
\nn\\
u_2  & = &  -\frac{7+3i\sqrt{L}-3L}{6\pi k_1}+ O(L^{-1/2})
\nn\\
u_3  & = &   \frac{4-3L}{12\pi k_1}+ O(L^{-1/2})\ .
\ee
In this case the anomalous dimension is predicted to be
\be
E_{\su(2)}^{(2)}(k_1) = \frac{8\pi^2}{L^3} k_1^2 (3L+7)+ O(L^{-4}) \qquad (k_1 = k_2,\ k_3=-2k_1)\ ,
\label{BAsu2NN}
\ee
which is again in agreement with the results of eqn.~(\ref{su2num})
and table~\ref{NUM_SU(2)_1Loop} (note that the fractional powers of
$L^{-1}$ have obligingly canceled out of the final expression for
the energy).

\subsection{Two and three-loop order}
A similar analysis can be performed on the two-loop $\su(2)$ spin-chain Hamiltonian. 
As before, we use the Jordan-Wigner transformation restricted to an odd-impurity
basis of operators to rewrite the two-loop Hamiltonian (\ref{Hsu2_2}) in terms of 
position-space fermionic oscillators, obtaining a result similar to eqn.~(\ref{Hsu2PS}):
\be
H_{\su(2)}^{(4)} & = & 
	\sum_{j=1}^L \biggl\{
	-\frac{1}{2}\Bigl[
	b_{j+2}^\dag b_{j} 
	+b_{j}^\dag b_{j+2}
	-4 \Bigl( b_{j+1}^\dag b_{j} 
	+ b_{j}^\dag b_{j+1}\Bigr)
	\Bigr]
	-3\,  b_{j}^\dag b_{j} 	
	-4\, b_{j}^\dag b_{j+1}^\dag b_{j} b_{j+1}
\nn\\
&&	+ b_{j+1}^\dag b_{j+2}^\dag b_{j} b_{j+1}
	+ b_{j}^\dag b_{j+1}^\dag b_{j+1} b_{j+2}
	+ b_{j}^\dag b_{j+2}^\dag b_{j} b_{j+2}
	\biggr\}\ .
\ee
Passing to momentum space, we obtain the two-loop analogue of eqn.~(\ref{Hsu2MS}):
\be
H_{\su(2)}^{(4)} & = & -8 \sum_{p=0}^{L-1} \sin^4\left(\frac{p\pi}{L}\right) \bt_p^\dag \bt_p
\nn\\
&&	+\frac{1}{L}\sum_{p,q,r,s=0}^{L-1} \left(
	e^{\frac{2\pi i (q+r)}{L}}
	+e^{\frac{-2\pi i (p+s)}{L}}
	+e^{\frac{4\pi i (q-s)}{L}}
	-4\,e^{\frac{2\pi i (q-s)}{L}}
	\right)
	 \bt_p^\dag \bt_q^\dag \bt_r \bt_s\, \delta_{p+q,r+s}\ .
\nn\\
&&
\label{Hsu2MS2}
\ee
Although the two-loop Hamiltonian includes ``long-range'' interactions among non-neighboring 
lattice sites, the momentum-space Hamiltonian (\ref{Hsu2MS2}) conveniently terminates
at two-body interaction terms. An equally important point is that, for fixed momenta
$p,q,\ldots$, the one-body (two-body) operators scale as $L^{-4}$ ($L^{-5}$) for large
$L$ (the corresponding scalings for the one-loop Hamiltonian were $L^{-2}$ ($L^{-3}$)).
This special relation between density scaling and power of coupling constant is critical
for matching to string theory.

We deal with the problem of finding the eigenvalues of the combined one- and two-loop 
Hamiltonian via Rayleigh-Schr\"odinger perturbation theory: at each value of the lattice 
length $L$ we treat the one-loop operator $H_{\su(2)}^{(2)}$ as a zeroth-order Hamiltonian 
and regard $H_{\su(2)}^{(4)}$ as a first-order perturbation. The $O(\lambda^2)$ corrections 
to the spectrum of $H_{\su(2)}^{(2)}$ are then found by taking expectation values of 
the perturbation $H_{\su(2)}^{(4)}$ in the (numerically-determined) eigenvectors of $H_{\su(2)}^{(2)}$. 
This is the recipe for non-degenerate first-order perturbation theory and we might worry
that the previously-noted parity-pair degeneracy of the eigenvalues of  $H_{\su(2)}^{(2)}$
would force us to use the rules of degenerate perturbation theory. As discussed in 
\cite{Beisert:2003tq,Callan:2004ev,Swanson:2004mk}, however, parity degeneracy can be traced to 
the existence of a higher Abelian charge which is conserved to at least three-loop 
order. This charge can be used to show that the formulas of non-degenerate perturbation 
theory can be used without modification. The basic observation is that conservation of
the Abelian charge guarantees that the matrix element of $H_{\su(2)}^{(4)}$ between two 
degenerate eigenstates of $H_{\su(2)}^{(2)}$ with different eigenvalues of the higher 
Abelian charge vanishes: this eliminates the vanishing energy-denominator singularities 
that would otherwise invalidate the non-degenerate first-order perturbation theory formulas 
(and similar arguments apply to the higher-order cases).

Using this method, we have evaluated the $O(\lambda^2)$ corrections to the spectrum
of anomalous dimensions for lattice sizes from $L=6$ to $L=40$. As before, we fit the 
spectral data to a power series in $1/L$ to read off the leading scaling
coefficients of the low-lying eigenvalues. As mentioned in the discussion of the
two-loop Hamiltonian (\ref{Hsu2MS2}), we expect the two-loop eigenvalues to have
the following scaling behavior in $1/L$:
\be
\label{scaling2}
E_{\alg L}^{(2)}(\lbrace k_i \rbrace) =  \frac{\lambda^2}{\alg L^4}{E^{(2,4)}
	(\lbrace k_i \rbrace)} + 
\frac{\lambda^2}{\alg L^5}{E^{(2,5)}(\lbrace k_i \rbrace)} + O(\lambda^2 \alg L^{-6})\ .
\ee
The numerical data confirm that the eigenvalues scale at least as fast as $L^{-4}$.
The resulting numerical values for the leading scaling coefficients of low-lying eigenvalues, 
$E_{\su(2)}^{(2,4)}$ and $E_{\su(2)}^{(2,5)}$, are presented in table~\ref{NUM_SU(2)_2Loop}. 
As originally reported in \cite{Callan:2004ev}, string theory makes the
following simple predictions for the two-loop large-$L$ expansion coefficients: 
\be
E_{\su(2)}^{(2,4)} = -(k_1^2+k_2^2+k_3^2)^2/16 & \qquad & k_1+k_2+k_3=0\nonumber\nonumber\\
E_{\su(2)}^{(2,5)}/E_{\su(2)}^{(2,3)}  =  8 & \qquad & (k_1\neq k_2\neq k_3)
\nonumber\\
E_{\su(2)}^{(2,5)}/E_{\su(2)}^{(2,3)} = \frac{76}{9} & \qquad &(k_1 = k_2,\ k_3=-2k_1)\ .
\label{su2num2}
\ee
The low-lying levels in the table match the string theory predictions quite accurately
and the decline in precision as one goes to higher energies is expected. As a consistency
check we note that this time we have no freedom to choose the momenta $(k_1,k_2,k_3)$
associated with each state: they have been fixed in the one-loop matching exercise.
\begin{table}[ht!]
\begin{eqnarray}
\begin{array}{|cccc|}
\hline
E_{\su(2)}^{(2,4)} & E_{\su(2)}^{(2,5)} & E_{\su(2)}^{(2,5)}/E_{\su(2)}^{(2,4)}  
                                  & (k_1,k_2,k_3) \\
\hline
-0.25-4.6\times 10^{-9}		&-2+8.0\times 10^{-7}	&8-3.4\times 10^{-6}	& (1,0,-1)	 \\
-2.25-1.4\times 10^{-6}		&-19+2.6\times 10^{-4}  &76/9+1.2\times 10^{-4} & (1,1,-2)	 \\	
-2.25-1.4\times 10^{-6}		&-19+2.6\times 10^{-4}	&76/9+1.2\times 10^{-4}	&  (-1,-1,2)	 \\
-4+8.3\times 10^{-7}		&-32-1.1\times 10^{-4}  &8+3.0\times 10^{-5}   &  (2,0,-2)	 \\
-12.25-9.9\times 10^{-6}	&-98+2.3\times 10^{-3}	&8-2.0\times 10^{-4}	&  (1,2,-3) 	 \\
-12.25-9.9\times 10^{-6}	&-98+2.3\times 10^{-3}	&8-2.0\times 10^{-4}	&  (-1,-2,3)	 \\
-20.25+3.2\times 10^{-3}	&-161.4			&7.97			&  (3,0,-3)	 \\
-36-2.8\times 10^{-3}		&-304.6			&8.46			&  (2,2,-4)	 \\
-36-2.8\times 10^{-3} 		&-304.6			&8.46			&  (-2,-2,4)	 \\
-42.25+4.9\times 10^{-3}	&-337.0			&7.97			&  (1,3,-4) 	 \\
-42.25+4.9\times 10^{-3}	&-337.0			&7.97			&  (-1,-3,4)	 \\
\hline
\end{array} \nonumber
\end{eqnarray}
\caption{Scaling limit of three-impurity
	$\su(2)$ numerical spectrum at two loops in $\lambda$}
\label{NUM_SU(2)_2Loop}
\end{table}


The three-loop $\su(2)$ Hamiltonian (\ref{Hsu2_3}) can be dealt with in a similar
fashion. The position space operator version of this Hamiltonian is too long 
to record here, but its momentum space version is fairly compact:
\be
H_{\su(2)}^{(6)} & = & 32 \sum_{p=0}^{L-1} \sin^6\left(\frac{p\pi}{L}\right)\bt_p^\dag \bt_p
	+\frac{1}{2L}\sum_{p,q,r,s=0}^{L-1}\biggl\{
	-10\, e^{\frac{2\pi i(q+r)}{L}} 
	+ e^{\frac{2\pi i(2q+r)}{L}}
	+ e^{\frac{2\pi i(q+2r)}{L}}
	+ e^{\frac{2\pi i(q-3s)}{L}}
\nn\\
&&	+ e^{\frac{2\pi i(2q-2r-3s)}{L}}
	+ e^{\frac{2\pi i(3q-2r-3s)}{L}}
	+ e^{\frac{2\pi i(q-r-3s)}{L}}
	+ e^{\frac{2\pi i(2q-r-3s)}{L}}
	- e^{\frac{2\pi i(q-2s)}{L}}
	-10\, e^{\frac{2\pi i(q-r-2s)}{L}}
\nn\\
&&	- e^{\frac{2\pi i(2q-r-2s)}{L}}
	- e^{\frac{2\pi i(3q-r-2s)}{L}}
	- e^{\frac{2\pi i(q+r-2s)}{L}}
	+29\, e^{\frac{2\pi i(q-s)}{L}}
	-10\, e^{\frac{4\pi i(q-s)}{L}}
	+ e^{\frac{6\pi i(q-s)}{L}}
\nn\\
&&	- e^{\frac{2\pi i(2q-s)}{L}}
	+ e^{\frac{2\pi i(3q-s)}{L}}
	- e^{\frac{2\pi i(q+r-s)}{L}}
	+ e^{\frac{2\pi i(2q+r-s)}{L}}
	+ e^{\frac{2\pi i(q+2r-s)}{L}}
	\biggr\}
	\bt_p^\dag \bt_q^\dag \bt_r\bt_s\,\delta_{p+q,r+s} 
\nn\\
&&	+ \frac{1}{L^2}\sum_{p,q,r,s,t,u=0}^{L-1}\biggl\{
	e^{\frac{2\pi i (q+3r-2t-3u)}{L}}
	+ e^{\frac{2\pi i (q+2r-s-2t-3u)}{L}}
\nn\\
&&\kern+80pt
	+ e^{\frac{2\pi i (2q+3r-t-3u)}{L}}
	+ e^{\frac{2\pi i (q+2r+s-u)}{L}}
	\biggr\}
	\bt_p^\dag \bt_q^\dag \bt_r^\dag
	\bt_s \bt_t \bt_u\, \delta_{p+q+r,s+t+u} \ .
\ee
It contains at most three-body operators and a careful examination of
terms shows that, for fixed momenta, the one-body operators scale as
$L^{-6}$, the two-body operators as $L^{-7}$ and so on. We therefore
expect the leading scaling coefficients in the $O(\lambda^3)$ eigenvalues
to be $E_{\su(2)}^{(3,6)}$ and $E_{\su(2)}^{(3,7)}$, to use a by-now-familiar
notation. To find the eigenvalues to this order, we continue with the
Rayleigh-Schr\"odinger perturbation theory strategy: the $O(\lambda^3)$
correction to any eigenvalue is the sum of the matrix element of $H_{\su(2)}^{(6)}$
in the appropriate eigenvector of $H_{\su(2)}^{(2)}$ plus the second-order
sum-over-states contribution of $H_{\su(2)}^{(4)}$. These two pieces
can easily be computed numerically from the explicit Hamiltonian operators
at a fixed $L$. Parity degeneracy and conservation of the higher Abelian
charge mentioned above continue to hold, and we can again use 
non-degenerate perturbation theory formulas to compute the eigenvalue 
corrections. We have generated numerical eigenvalue data for lattices
from $L=6$ to $L=40$ and the large-$L$ scaling coefficients of the low-lying
states extracted from those data are given in table~\ref{NUM_SU(2)_3Loop}. 
\begin{table}[ht!]
\begin{eqnarray}
\begin{array}{|cccc|}
\hline
E_{\su(2)}^{(3,4)} & E_{\su(2)}^{(3,7)} & E_{\su(2)}^{(3,7)}/E_{\su(2)}^{(3,6)}  
                                  & (k_1,k_2,k_3) \\
\hline
0.1250		&2.0003		&16.003		& (1,0,-1)	 \\
4.125		&58.03  	&14.07 		& (1,1,-2)	 \\	
4.125		&58.03		&14.07		&  (-1,-1,2)	 \\
7.999		&128.2		&16.03		&  (2,0,-2)	 \\
49.62		&713.3		&14.37		&  (1,2,-3) 	 \\
49.62		&713.3		&14.37		&  (-1,-2,3)	 \\
91.15		&1,454		&15.96		&  (3,0,-3)	 \\
263.8		&3,739		&14.17		&  (2,2,-4)	 \\
263.8 		&3,739		&14.17		&  (-2,-2,4)	 \\
\hline
\end{array} \nonumber
\end{eqnarray}
\caption{Scaling limit of three-impurity
	$\su(2)$ numerical spectrum at three loops in $\lambda$}
\label{NUM_SU(2)_3Loop}
\end{table}
As is by now well-known, the detailed match to string theory breaks down at 
three-loop order, so there is no point in trying to match these results to string 
predictions. We were initially motivated to pursue a virial treatment of
this problem because the more classic Bethe ansatz methods were not yet able
to deal with non-local spin chains. Important
progress has recently been made on the Bethe ansatz side and it may be useful
to compare what can be done by each method (results should of course agree).

A modified Bethe ansatz for the $\su(2)$ sector of the gauge theory, possibly 
incorporating all orders of higher-loop physics, has recently been proposed 
\cite{Serban:2004jf,Beisert:2004hm}.\footnote{The long-range ansatz based on the 
Inozemtsev spin chain in \cite{Serban:2004jf} suffers from improper 
BMN scaling at four-loop order, a problem that is surmounted in \cite{Beisert:2004hm}. }
It is an instructive exercise, and a useful consistency check on this bold proposal, 
to verify that it reproduces the higher-loop scaling coefficients for three 
impurity anomalous dimensions that we have just computed by virial methods 
(and displayed in tables ~\ref{NUM_SU(2)_1Loop},~\ref{NUM_SU(2)_2Loop} 
and \ref{NUM_SU(2)_3Loop}). For completeness, we briefly summarize the
new ansatz, referring the reader to \cite{Beisert:2004hm} for a detailed account.
In the new ansatz, the momenta $p_i$ of the excitations (closely related to the 
Bethe roots) become functions of $\lambda$ (as well as $L$ 
and mode numbers) and are determined by a modified version of eqns.~(\ref{BAESU2_1},\ref{BAESU2_2}):
\be
e^{i L p_i} & = &\prod_{j\neq i}^I 
	\frac{\varphi(p_i)-\varphi(p_j)+i}{\varphi(p_i)-\varphi(p_j)-i} \qquad 
\sum_{i=1}^I p_i = 0\ .
\label{LRBA}
\ee
Dependence on $\lambda$ 
enters through the phase function $\varphi(p_i)$ which is defined in terms of the 
excitation momenta $p_i$ as follows:
\be
\varphi(p_i) & \equiv & \frac{1}{2}\cot\left({p_i}/{2}\right)
	\sqrt{1+\frac{\lambda}{\pi^2}\sin^2\left({p_i}/{2}\right)}\ .
\ee
The energy eigenvalue corresponding to a particular root of these equations is given in 
terms of the excitation momenta $p_i$ by the formula
\be
E_{\su(2)} = \sum_{i=1}^I \frac{8\pi^2}{\lambda}\left(
	\sqrt{1+\frac{\lambda}{\pi^2}\sin^2 \left(p_i/2\right)} -1\right)\ .
\ee
Finding exact solutions of these equations is even more difficult than before,
but we can follow the previous strategy of developing an expansion in powers of 
$1/L$ about non-interacting impurities on an infinite lattice. This is achieved 
by expanding the excitation momenta $p_i$ according to
\be
p_i = \frac{2\pi k_i}{L} + \sum_{n=1}\frac{p_i^{(n)}}{L^{\frac{n+2}{2}}}\ ,
\ee
where the integers $k_i$ (subject to the cyclicity constraint $\sum_i k_i = 0$) 
characterize the non-interacting state about which the expansion is developed.
The appearance of half-integer powers of $L^{-1}$ in this expansion is needed
to accommodate bound-state solutions to the Bethe equations which arise when
some of the momenta $k_i$ are equal. Solutions to the Bethe equation (\ref{LRBA}) 
will determine the expansion coefficients $p_{i}^{(n)}$ in terms of the mode 
numbers $k_i$ and ultimately lead to expansions of the energies as power 
series in $L^{-1}$, with coefficients that are functions of $\lambda/L^2$.

Explicit results for the $L^{-1}$ expansion of gauge theory operators of arbitrary 
impurity number, derived by the above method, were presented in 
\cite{Arutyunov:2004vx}.\footnote{It is important to note that the focus of this
paper is a \emph{different} Bethe ansatz, designed to match the spectrum of
the string theory: the gauge theory Bethe ansatz results are derived for comparison
purposes.}
As usual, expressions are different depending on whether all momenta are unequal or some
subset of them are equal. For all mode numbers $k_i$ unequal the $I$-impurity energy formula 
in \cite{Arutyunov:2004vx} is
\be
\label{abanearBMN}
E_{\su(2)} & = & L-I+\sum_{i=1}^I
	\left(\sqrt{1+\lambda'\,k_i^2}
	-\frac{\lambda'}{L-I} \frac{I\, k_i^2}
	{\sqrt{1+\lambda'\,k_i^2}} 
	\right)
\nn\\
&&	-\frac{\lambda'}{L-I}
	\sum_{\textstyle\atopfrac{i,j=1}{i\neq j}}^I \frac{2k^2_i k_j}{k^2_i-k^2_j}
	\left(
	k_j+k_i\sqrt{\frac{1+\lambda'\,k_j^2}{1+\lambda'\,k_i^2}} \right)
		+O(L^{-2})\ ,
\label{LRBAeng}
\ee
where we have used $\lambda'=\lambda/J^2 = \lambda/(L-I)^2$ for convenience 
($J=L-I$ is the total $R$-charge). 
To compare with our virial results, we must further expand in
$\lambda$; expanding to first and second order yields the following scaling coefficients 
(valid for all $k_i$ unequal):
\be
E_{\su(2)}^{(1,2)} = k_1^2 + k_1k_2 + k_2^2 & \qquad &
E_{\su(2)}^{(1,3)} = 2(k_1^2 + k_1k_2 + k_2^2) \nn\\
E_{\su(2)}^{(2,4)} = -\frac{1}{4}(q^2 + qr + r^2)^2 &\qquad&
E_{\su(2)}^{(2,5)} = -2(q^2 + qr + r^2)^2\ . 
\ee
These one- and two-loop coefficients match the numerical results presented in 
tables~\ref{NUM_SU(2)_1Loop} and \ref{NUM_SU(2)_2Loop} and the analytic string
formulas in eqns.~(\ref{su2num},\ref{su2num2}). It is harder to
write down a general formula for the many cases in which subsets of momenta are
equal but the solution for the particular case of three impurities with a two-excitation 
bound state $(k_1 = k_2 = n,\ k_3=-2n)$ was also presented in \cite{Arutyunov:2004vx}:
\be
E_{\su(2)}&=&L-3+2 \sqrt{1+\lambda'\,n^2}+\sqrt{1+\lambda'\,4 n^2} 
\nn\\
&&\kern-50pt	
	-\frac{\lambda'\,n^2}{L-3}
	\left(
	\frac{1}{1 + \lambda'\,n^2} + \frac{6}{{\sqrt{1 + \lambda'\,n^2}}} +
	\frac{12}{{\sqrt{1 + \lambda'\,4 n^2}}} -
  	\frac{8}{{\sqrt{1 + \lambda'\,n^2}}\,{\sqrt{1 + \lambda'\,4 n^2}}}
	\right)\ .
\label{LRBAnn}
\ee
To compare with the virial results, one must again expand the energy in powers of 
$\lambda$. Doing so yields the following one- and two-loop bound-state scaling coefficients:
\be
E_{\su(2)}^{(1,2)} = 3\,n^2 & \qquad &
E_{\su(2)}^{(1,3)} = 7\,n^2 \nn\\
E_{\su(2)}^{(2,4)} = -\frac{9}{4}\,n^4 &\qquad&
E_{\su(2)}^{(2,5)} = -19\,n^4\ .
\ee 
We easily verify that this agrees with numerical virial results to two-loop order.

The three-loop coefficients obtained by expanding the energy formulas
in eqns.~(\ref{LRBAeng},\ref{LRBAnn}) are given by
\be
E_{\su(2)}^{(3,6)} & = & \frac{1}{16}\left(
	2\,{{k_1}}^6 + 6\,{{k_1}}^5\,{k_2} + 15\,{{k_1}}^4\,{{k_2}}^2 + 
    20\,{{k_1}}^3\,{{k_2}}^3 + 15\,{{k_1}}^2\,{{k_2}}^4 + 6\,{k_1}\,{{k_2}}^5 + 
    2\,{{k_2}}^6\right)
\nn\\
E_{\su(2)}^{(3,7)} & = &
	\frac{1}{4}\left(
	8\,{{k_1}}^6 + 24\,{{k_1}}^5\,{k_2} + 51\,{{k_1}}^4\,{{k_2}}^2 + 
    62\,{{k_1}}^3\,{{k_2}}^3 + 51\,{{k_1}}^2\,{{k_2}}^4 + 24\,{k_1}\,{{k_2}}^5 + 
    8\,{{k_2}}^6\right)\ ,
\nn\\
&&
\ee
for $(k_1\neq k_2\neq k_3)$, and 
\be
E_{\su(2)}^{(3,6)} = \frac{33}{8}\,n^6 \qquad E_{\su(2)}^{(3,7)} = 58\,n^6\ ,
\ee
for the bound-state solution with $(k_1=k_2=n,\ k_3=-2n)$. The numerical values 
of these $O(\lambda^3)$ coefficients are tabulated for several low-lying states 
in the spectrum in table~\ref{BA3}. The correspondence with table~\ref{NUM_SU(2)_3Loop}, 
which displays the three-loop expansion coefficients extracted from numerical
diagonalization of the three-loop Hamiltonian, is good.  At this order in the loop 
expansion higher-order $1/L$ corrections to the spectrum are more important 
(compared to the one- and two-loop cases), and the numerical extrapolation is
less reliable (especially as the lattice momenta increase). The precision can always 
be improved by including data from larger lattices in the extrapolation.  
We emphasize that this discussion concerns the different methods of calculation 
of operator dimensions in the $\su(2)$ sector only. It seems to
us to give useful further evidence that the long-range Bethe ansatz for the $\su(2)$
sector of the gauge theory \cite{Beisert:2004hm} is exact. 
\begin{table}[ht!]
\begin{eqnarray}
\begin{array}{|cccc|}
\hline
E_{\su(2)}^{(3,6)} & E_{\su(2)}^{(3,7)} &E_{\su(2)}^{(3,7)}/E_{\su(2)}^{(3,6)}
					&  (k_1,k_2,k_3) \\
\hline
0.125	& 2	& 16 		& (1,0,-1)	 \\
4.125	& 58	& 14.06		& (1,1,-2)	\\
4.125	& 58	& 14.06		& (-1,-1,2)	\\
8	& 128 	& 16		& (2,0,-2)	 \\
49.625  & 713   & 14.37  	& (1,2,-3)	 \\
49.625  & 713   & 14.37  	& (-1,-2,3)	 \\
91.125  & 1,458 & 16		& (3,0,-3)	\\
264	& 3,712	& 14.06		& (2,2,-4)	\\
264	& 3,712	& 14.06		& (-2,-2,4)	\\
\hline
\end{array} \nonumber
\end{eqnarray}
\caption{Three-impurity $\su(2)$ spectrum from the long-range 
	Bethe ansatz at three loops}
\label{BA3}
\end{table}

\section{A closed $\su(1|1)$ subsector of $\su(2|3)$ }
The three-impurity string theory analysis of \cite{Callan:2004ev} identified 
a fermionic sector of the theory which is diagonalized by string states composed 
of fermionic excitations projected onto particular four-dimensional subspaces 
(which transform in an $SU(2)^2\times SU(2)^2$ notation 
as a $({\bf 2,1;2,1})$ or $({\bf 1,2;1,2})$ of $SO(4)\times SO(4)$) 
and symmetrized in their $SO(4)\times SO(4)$ indices. 
It was also shown that this three-impurity 
subsector of the theory decouples at all orders in $\lambda$.  

On the gauge theory side this subsector corresponds to an $\su(1|1)$ 
subgroup of the closed $\su(2|3)$ sector studied by Beisert in 
\cite{Beisert:2003ys,Beisert:2003jj}.
(Supersymmetric integrable $\su(n|m)$ spin chains have previously been studied in 
certain condensed-matter applications; see, eg.,~\cite{liuwang}.)
In the present setting the fields of $\su(2|3)$ consist of three complex scalars $\phi_a$ and 
two complex fermions $\psi_\alpha$.  In the closed $\su(1|1)$ subspace we restrict
to a single scalar denoted by $Z$ and a single fermion labeled by $\psi$.
Just as in the $\su(2)$ sector, we use the fermionic position-space oscillators
$b_j^\dag,\ b_j$ to create or annihilate fermionic $\psi$ insertions in a ground
state composed of $L$ scalars:
\be
\Ket{L} = \tr(Z^L) \qquad 
b_j^\dag \ket{L} = \tr( Z_1\cdots Z_{j-1} \psi Z_{j+1} \cdots Z_L )\ .
\label{oscsu11}
\ee

In \cite{Beisert:2003ys}, Beisert gave the action of the Hamiltonian on the $\su(2|3)$ 
spin chain to three-loop order.\footnote{Beisert's three-loop Hamiltonian was restricted
in \cite{Beisert:2003ys} to the bosonic sector, but the author has since provided us with 
the complete version.}  In the notation of \cite{Beisert:2003ys}, the action of the Hamiltonian on basis 
states can be represented in terms of special permutation operators denoted by
\be
\genfrac{\{\}}{}{0pt}{0}{A_1 \ldots A_N }{ B_1 \ldots B_N }\ ,
\nn
\ee
which replace all occurrences of the upper sequence of fields 
$A_1\ldots A_N$ in the trace by the lower sequence $B_1\ldots B_N$.  
Restricting Beisert's $\su(2|3)$ Hamiltonian to the $\su(1|1)$ subsector 
at one-loop order yields
\be
H_{\su(1|1)}^{(2)} =  
	\genfrac{\{\}}{}{0pt}{0}{Z\psi }{ Z\psi  } 
	+  \genfrac{\{\}}{}{0pt}{0}{\psi Z }{ \psi Z}
	-  \genfrac{\{\}}{}{0pt}{0}{Z\psi }{ \psi Z} 
	-  \genfrac{\{\}}{}{0pt}{0}{\psi Z }{ Z\psi}
	+ 2  \genfrac{\{\}}{}{0pt}{0}{\psi\psi }{ \psi\psi}\ .
\label{BsrtH2}
\ee
In terms of the position-space oscillators of eqn.~(\ref{oscsu11}), the $\su(1|1)$
Hamiltonian can be assembled by inspection and takes the form
\be
H_{\su(1|1)}^{(2)} = \sum_{j=1}^L\left(
	b_j^\dag b_j + b_{j+1}^\dag b_{j+1} - b_{j+1}^\dag b_j - b_j^\dag b_{j+1}
	\right)\ .
\ee
There are no higher-body interaction terms at
this order in $\lambda$.  This fact can be checked by computing 
\be
\braket{ L| b_{i+1}b_i (H_{\su(1|1)}^{(2)}) b_i^\dag b_{i+1}^\dag | L} = 2\ ,
\ee
which reproduces the two-body matrix element given by the last term
in eqn.~(\ref{BsrtH2}).  In momentum space we obtain
\be
H_{\su(1|1)}^{(2)} = 4\sum_{p=0}^{L-1}\sin^2\left(\frac{p\pi}{L}\right)
	\bt_p^\dag \bt_p\ .
\label{su11oneloop}
\ee

The two-loop $\su(1|1)$ momentum-space Hamiltonian can be extracted in the same manner
(the position-space version is too long to print here):
\be
H_{\su(1|1)}^{(4)} & = & -8\sum_{p=0}^{L-1}\sin^4\left(\frac{p\pi}{L}\right)
	\bt_p^\dag \bt_p
	+\frac{1}{4L}\sum_{p,q,r,s=0}^{L-1}\biggl\{
	e^{\frac{2\pi i (q-2r)}{L}}
	+ e^{\frac{2\pi i (2q-r)}{L}}
	-4\, e^{\frac{2\pi i (q-r)}{L}}
\nn\\
&&\kern-20pt
	-2\, e^{\frac{2\pi i (q-2r-s)}{L}}
	-2\, e^{\frac{2\pi i (q+s)}{L}}
	+ e^{\frac{2\pi i (q-r+s)}{L}}
	+ e^{\frac{2\pi i (2q-2r-s)}{L}}
	\biggr\} \bt_p^\dag \bt_q^\dag \bt_r \bt_s\, \delta_{p+q,r+s}\ .
\ee
Finally, the complete three-loop Hamiltonian for this subsector is
\be
H_{\su(1|1)}^{(6)} & = & 32\sum_{p=0}^{L-1}\sin^6\left(\frac{p\pi}{L}\right)
	\bt_p^\dag \bt_p
	-\frac{1}{16}\sum_{p,q,r,s=0}^{L-1}e^{\frac{60\pi i(q-r)}{L}}\biggl\{
	2\, e^{-\frac{2\pi i (27q-29r)}{L}}
	+2\, e^{-\frac{2\pi i (28q-29r)}{L}}
\nn\\
&&	-4\, e^{-\frac{2\pi i (27q-28r)}{L}}
	+37\, e^{-\frac{2\pi i (29q-28r)}{L}}
	-6\, e^{-\frac{2\pi i (29q-27r)}{L}}
	+8\, e^{-\frac{56\pi i (q-r)}{L}}
	-72\, e^{-\frac{58\pi i (q-r)}{L}}
\nn\\
&&	-6\, e^{-\frac{2\pi i (29q-29r-2s)}{L}}
	-40\, e^{-\frac{2\pi i (29q-30r-s)}{L}}
	+37\, e^{-\frac{2\pi i (29q-29r-s)}{L}}
	-8\, e^{-\frac{2\pi i (29q-28r-s)}{L}}
\nn\\
&&	+8\, e^{-\frac{2\pi i (27q-28r+s)}{L}}
	+2\, e^{-\frac{2\pi i (28q-28r+s)}{L}}
	-40\, e^{-\frac{2\pi i (29q-28r+s)}{L}}
	-4\, e^{-\frac{2\pi i (27q-27r+s)}{L}}
\nn\\
&&	+8\, e^{-\frac{2\pi i (29q-27r+s)}{L}}
	+2\, e^{-\frac{2\pi i (27q-27r+2s)}{L}}
	+8\, e^{-\frac{2\pi i (29q-30r-2s)}{L}}
	\biggr\}\bt_p^\dag \bt_q^\dag \bt_r \bt_s\, \delta_{p+q,r+s}
\nn\\
&&	+\frac{1}{16}\sum_{p,q,r,s,t,u=0}^{L-1}\biggl\{
	2\, e^{\frac{2\pi i (q+2r-3s-2t )}{L}}
	- e^{\frac{2\pi i (q+3r-3s-2t )}{L}}
	-4\, e^{\frac{2\pi i (q+2r-3s-t )}{L}}
\nn\\
&&	- e^{\frac{2\pi i (2q+3r-3s-t )}{L}}
	+8\, e^{\frac{2\pi i (q+2r-2s-t )}{L}}
	+2\, e^{\frac{2\pi i (2q+3r-2s-t )}{L}}
	-4\, e^{\frac{2\pi i (q+2r-3s-2t-u )}{L}}
\nn\\
&&	+2\, e^{\frac{2\pi i (q+3r-3s-2t-u )}{L}}
	+2\, e^{\frac{2\pi i (q+2r-2s+u )}{L}}
\nn\\
&&\kern+40pt
	-4\, e^{\frac{2\pi i (q+2r-s+u )}{L}}
	-4\, e^{\frac{2\pi i (q+2r-2s-t+u )}{L}}
	\biggr\} \bt_p^\dag \bt_q^\dag \bt_r^\dag \bt_s \bt_t \bt_u\,
	\delta_{p+q+r,s+t+u}\ .
\ee
We note that $H_{\su(1|1)}^{(2)}$, $H_{\su(1|1)}^{(4)}$ and $H_{\su(1|1)}^{(6)}$
terminate at one-body, two-body and three-body interactions, respectively.
This will permit us to obtain the exact $L$-dependence of successive 
terms in the $\lambda$ expansion of energy eigenvalues.

As in the $\su(2)$ sector, we can use non-degenerate perturbation theory 
to extract the $L^{-1}$ scaling coefficients of the $\su(1|1)$ eigenvalue spectrum 
up to three loops in $\lambda$. The scaling coefficients extrapolated from
numerical diagonalization of lattices up to $L=40$ are recorded  
for one-loop, two-loop and three-loop orders in tables~\ref{NUM_fermi_1Loop}, 
\ref{NUM_fermi_2Loop}, and \ref{NUM_fermi_3Loop}, respectively. The same increase 
in leading power of $L^{-1}$ with corresponding order in $\lambda$ 
that was noted in the $\su(2)$ sector is found here as well 
(we use the same notation for the scaling coefficients as before in order to keep 
track of these powers). It should also be noted 
that, because the impurities in this sector are fermions symmetrized on all
group indices, the lattice momenta of all pseudoparticles must be different.
The string theory results of \cite{Callan:2004ev} amount to the following
predictions for the one-loop and two-loop scaling coefficients:
\be
E_{\su(1|1)}^{(1,2)} = (k_1^2+k_1k_2+k_2^2) &\qquad& E_{\su(1|1)}^{(1,3)} = 0 \nn\\
E_{\su(1|1)}^{(2,4)} -\frac{1}{4}(k_1^2+k_1k_2+k_2^2)^2  &\qquad&  
E_{\su(1|1)}^{(2,5)} = -(k_1^2+k_1k_2+k_2^2)^2 \ .
\label{su11ext}
\ee
The agreement of these predictions with the data in tables~\ref{NUM_fermi_1Loop} 
and \ref{NUM_fermi_2Loop} is excellent (with the usual caveat that data
on larger and larger lattices is required to maintain a fixed precision as
one goes to higher and higher energy levels).
\begin{table}[ht!]
\begin{eqnarray}
\begin{array}{|cccc|}
\hline
E_{\su(1|1)}^{(1,2)} & E_{\su(1|1)}^{(1,3)} & E_{\su(1|1)}^{(1,3)}/E_{\su(1|1)}^{(1,2)}  
		& (k_1,k_2,k_3)  \\
\hline
1+1.3 \times 10^{-10} 	&	-1.9\times 10^{-8}	& -1.9\times 10^{-8} 	&(1,0,-1) \\
4-1.0\times 10^{-7} 	&	1.8\times 10^{-5}	& 4.6\times 10^{-6}	&	(2,0,-2)  \\
7-2.5\times 10^{-7}  	&	4.4\times 10^{-5}	  & 6.3\times 10^{-6} 	&	(1,2,-3)  \\
7-2.5\times 10^{-7}  	&	4.4\times 10^{-5} 	  & 6.3\times 10^{-6}	&	(-1,-2,3)  \\
9-3.9\times 10^{-7} 	&	7.9\times 10^{-5}	   & 8.7\times 10^{-6}	&	(3,0,-3)  \\
13-4.0\times 10^{-6}  	&	8.2\times 10^{-4}	  & 6.3 \times 10^{-5}	&	(1,3,-4)    \\
13-4.0\times 10^{-6}  	&	8.2\times 10^{-4}	   & 6.3 \times 10^{-5} &	(-1,-3,4)    \\
16-2.0\times 10^{-5}  	&	4.1\times 10^{-3}	  & 2.6 \times 10^{-4}	&	(4,0,-4)      \\
19-3.5\times 10^{-5}  	&	7.3\times 10^{-3}	  & 3.8 \times 10^{-4}	&	(2,3,-5)    \\
19-3.5\times 10^{-5}  	&	7.3\times 10^{-3}	  & 3.8 \times 10^{-4}	&	(-2,-3,5)    \\
\hline
\end{array} \nonumber
\end{eqnarray}
\caption{Scaling limit of one-loop numerical spectrum of three-impurity 
	$\su(1|1)$ subsector}
\label{NUM_fermi_1Loop}
\end{table}
\begin{table}[ht!]
\begin{eqnarray}
\begin{array}{|cccc|}
\hline
E_{\su(1|1)}^{(2,4)} & E_{\su(1|1)}^{(2,5)} & E_{\su(1|1)}^{(2,5)}/E_{\su(1|1)}^{(2,4)}  
		& (k_1,k_2,k_3)   \\
\hline
-0.25	  	  &-0.99999 &	  	3.99995  & (1,0,-1)   \\
-4.00006	  &-15.990 & 		3.998  &(2,0,-2)    \\	
-12.251		  &-48.899 &	  	3.992  &(1,2,-3)    \\
-12.251		  &-48.899 &	  	3.992  &(-1,-2,3)    \\
-20.25		  &-80.89 &	  	3.995   & (3,0,-3)   \\
-42.25		  &-168.2 &	  	3.98  &(1,3,-4)    \\
-42.25		  &-168.2 &	  	3.98  &(-1,-3,4)    \\
-64.00		  &-254.6 & 	  	3.98  &(4,0,-4)    \\
-90.26		  &-359.3 &	  	3.98   &(2,3,-5)    \\
-90.26		  &-359.8 &  		3.99  &(-2,-3,5)   \\
\hline
\end{array} \nonumber
\end{eqnarray}
\caption{Scaling limit of two-loop numerical spectrum of three-impurity 
	$\su(1|1)$ subsector }
\label{NUM_fermi_2Loop}
\end{table}

The scaling limit of the three-loop ratio $E_{\su(1|1)}^{(3,7)}/E_{\su(1|1)}^{(3,6)}$
is recorded for the first few low-lying states in the spectrum in 
table~\ref{NUM_fermi_3Loop}. These values are in disagreement with the corresponding 
three-loop predictions from the string theory as can be seen by comparing with
the results of \cite{Callan:2004ev}. Given the well-established three-loop 
disagreement between the string and gauge theory in the $\su(2)$ sector, however, 
this disagreement in the $\su(1|1)$ subsector is not unexpected. 
\begin{table}[ht!]
\begin{eqnarray}
\begin{array}{|cc|}
\hline
 E_{\su(1|1)}^{(3,7)}/E_{\su(1|1)}^{(3,6)}  
		& (k_1,k_2,k_3)   \\
\hline
  		-86.41  & (1,0,-1) \\
		-85.71  &(2,0,-2) \\	
	  	-83.74  &(1,2,-3) \\
	  	-83.74  &(-1,-2,3) \\
	  	-101.9   & (3,0,-3) \\
	  	-96.01  &(1,3,-4) \\
	  	-96.01  &(-1,-3,4) \\
	  	-158.1 &(4,0,-4) \\
\hline
\end{array} \nonumber
\end{eqnarray}
\caption{Scaling limit of three-loop numerical spectrum of three-impurity 
	$\su(1|1)$ fermionic subsector }
\label{NUM_fermi_3Loop}
\end{table}

The extrapolated gauge theory results in eqn.~(\ref{su11ext}) for the one-loop 
coefficients $E_{\su(1|1)}^{(1,3)}$ and $E_{\su(1|1)}^{(1,2)}$ should be checked 
against the predictions of the general one-loop Bethe ansatz \cite{Minahan:2002ve,Beisert:2003yb}
applied to the $\su(1|1)$ sector (as far as we know, no higher-loop Bethe
ansatz is available here). To apply the general Bethe ansatz equation of 
eqn.~(\ref{BAEFull}), 
we note that the $\su(1|1)$ Dynkin diagram is just a single fermionic node: 
the Cartan matrix is empty and the single Dynkin label is $V_{\su(1|1)}=1$ \cite{LSA1,LSA2}.
We therefore obtain the simple one-loop Bethe equation
\be
\left( \frac{u_{i} + \frac{i}{2} }{u_{i}-\frac{i}{2}}\right)^L = 1\ .
\label{BAEFull2}
\ee
Rather remarkably, eqn.~(\ref{BAEFull2}) 
can be solved exactly for arbitrary impurity number! The general $\su(1|1)$ 
Bethe roots are
\be
u_{i} = \frac{1}{2}\cot \left(\frac{k_i \pi}{L}\right)
\ee
and the energy eigenvalues computed from eqn.~(\ref{betheenergy1L}) are
\be
E_{\su(1|1)} = 4\sum_{i=1}^I\sin^2\left(\frac{\pi k_i}{L}\right)\ ,
\label{BAsu11}
\ee
with the usual condition $\sum k_i = 0 \mod L$ from eqn.~(\ref{BAE2}).  
This is just the sum of free lattice Laplacian energies and clearly matches 
the energies one would obtain from the one-loop $\su(1|1)$ Hamiltonian 
of eqn.~(\ref{su11oneloop}) (since the latter has no interaction terms). 
No expansion in $1/L$ was necessary in this argument, but it is straightforward to
expand the energies in $1/L$ and verify the numerical results obtained 
in table~\ref{NUM_fermi_1Loop} and eqn.~(\ref{su11ext}).

\section{The $\Sl(2)$ sector}
As noted in \cite{Callan:2004ev}, integrable $\Sl(2)$ spin chains have previously
been the subject of several studies involving, among other interesting problems, 
high-energy scattering amplitudes in non-supersymmetric QCD 
(see, eg.,~\cite{Belitsky:2003ys} and references therein).
The $\Sl(2)$ closed sector of ${\cal N}=4$ SYM was studied in \cite{Beisert:2003jj}, and
the spin chain Hamiltonian in this sector is presently known to one loop in $\lambda$.

The constituent fields in this sector are $SO(6)$ bosons $Z$ carrying a single unit
of $R$-charge ($Z = \phi_5 + i\phi_6$),
and each lattice site on the $\Sl(2)$ spin chain is occupied 
by a single $Z$ field acted on by any number of the spacetime covariant derivatives
${\cal D}\equiv {\cal D}_1+i{\cal D}_2$.  
The total $R$-charge of a particular operator is therefore equal to the 
lattice length $L$, and an $I$-impurity operator basis is spanned by single-trace
operators carrying all possible distributions of $I$ derivatives among the
$L$ lattice sites:
\be
{\rm Tr}\left({\cal D}^IZ~Z^{{\alg L}	-1} \right)\ ,\quad  
{\rm Tr}\left({\cal D}^{I-1}Z~{\cal D}Z~Z^{{\alg L}	-2} \right)\ ,\quad
	~{\rm Tr}\left({\cal D}^{I-1}Z~Z{\cal D}Z~Z^{{\alg L}	-3} \right)\ ,\quad \ldots
\ee
The integer $I$ counts the total number of derivatives in the operator and, since any 
number of impurities can occupy the same lattice site, one can think of $n$ derivative 
insertions at the $i^{th}$ lattice site as $n$ bosonic oscillator excitations at 
the $i^{th}$ lattice position:
\be
(a_i^\dag)^n \ket{\alg L} \sim {\rm Tr} \left( Z^{i-1} {\cal D}^n Z  Z^{{\alg L}-i} \right)\ ,
\ \ldots
\ee
The ground state $\ket{L}$ is represented by a length $L$ chain with 
no derivative insertions: $\ket{\alg L} = {\rm Tr} \left(Z^{\alg L} \right)$.

The one-loop $\Sl(2)$ spin chain Hamiltonian (corresponding to the dilatation 
operator in this sector) was constructed in \cite{Beisert:2003jj} and was defined 
by its action on basis states rather than directly expressed as an operator:
\be
&&H_{\Sl(2)}^{(2)}  = 
	\sum_{j=1}^{\alg L} H^{\Sl(2)}_{j,j+1}\ , 
\nn\\
&&H^{\Sl(2)}_{1,2} (a_1^\dag)^j (a_2^\dag)^{n-j}\ket{\alg L}  =  
	\sum_{j'=0}^{n}\left[
	\delta_{j=j'}\left(h(j)+h(n-j)\right)
	-\frac{\delta_{j\neq j'}}{|j-j'|}\right]
	(a_1^\dag)^{j'}(a_2^\dag)^{n-j'}\ket{{\alg L}}\ 
\nn\\
&&
\label{sl2int}
\ee
(where $h(n)=1+\ldots+1/n$ are the harmonic numbers).
In other words, $H_{\Sl(2)}^{(2)}$ is a sum over the position-space Hamiltonian
$H_{j,j+1}^{\Sl(2)}$ which acts on the $j^{th}$ and $(j+1)^{th}$ (neighboring)
lattice sites; the action of $H_{j,j+1}^{\Sl(2)}$ can be summarized by the
explicit form given for $H_{1,2}^{\Sl(2)}$ above.  Since it
is only defined by its action on the state $(a_1^\dag)^j (a_2^\dag)^{n-j}\ket{\alg L}$,
it is difficult to immediately translate $H_{\Sl(2)}^{(2)}$ to momentum space.
However, it is possible to expand it in powers of fields and use eqn.~(\ref{sl2int})
to iteratively determine the expansion coefficients. The virial argument 
furthermore tells us that higher powers in the fields will determine 
higher powers of $L^{-1}$ in the expansion of the energy. For our current
purposes, it suffices to know the Hamiltonian expanded out to terms of
fourth order in the fields and this truncation of the 
Hamiltonian can easily be constructed by inspection:
\be
H_{\Sl(2)}^{(2)} & = & -\sum_{j=1}^L \left[
	\Bigl(a_{j+1}^\dag-2a_j^\dag+a_{j-1}^\dag\Bigr)
	\Bigl(a_j - \frac{1}{2}a_j^\dag a_j^2 \Bigr)
	+\frac{1}{4}\Bigl({a_{j+1}^{\dag\ 2}}-2{a_{j}^{\dag\ 2}}+{a_{j-1}^{\dag\ 2}}\Bigr)
	a_j^2 \right]\ + \cdots
\nn\\
&&
\ee
Transformation to momentum space gives
\begin{eqnarray}
&&\kern-25pt	H_{\Sl(2)}^{(2)}  = 
	~{\sum_{p=0}^{L-1}}4\sin^{2}\frac{p \pi}{\alg L}~\at_{p}^{\dag}\at_{p} \nn\\
&&\kern+15pt 	+ \frac{1}{\alg L} {\sum_{p,q,r,s=0}^{L-1}}\delta_{p+q,r+s}
		\left(-\sin^{2}\frac{p\pi}{\alg L}
		-\sin^{2}\frac{q\pi}{\alg L}+\sin^{2}\frac{(p+q)\pi}{\alg L}\right)  
		\at_{p}^{\dag}\at_{q}^{\dag }\at_{r}\at_{s}~
		+\cdots
\label{hamilsltwofin}
\end{eqnarray}
This Hamiltonian acts on an $I$-impurity Fock space spanned by the generic states
\be
\at_{k_1}^\dag \at_{k_2}^\dag \at_{k_3}^\dag \cdots \ket{L}\ ,
\ee
with lattice momenta labeled by 
$k_i = 0,\dots,L-1$, and subject to the constraint $\sum_{i} k_i = 0 \mod L$.  
Numerically diagonalizing this Hamiltonian on a range of lattice sizes, we obtain
data from which we extract the numerical predictions for the one-loop coefficients 
$E_{\Sl(2)}^{(1,2)}$ and $E_{\Sl(2)}^{(1,3)}$ presented in table~\ref{NUM_SL(2)_1Loop}.
String theory makes the following predictions \cite{Callan:2004ev}
for the scaling coefficients 
\be
E_{\Sl(2)}^{(1,2)} = (k_1^2 +k_1 k_2+ k_2^2 ) \qquad
E_{\Sl(2)}^{(1,3)}/E_{\Sl(2)}^{(1,2)} = -2 \qquad k_1\neq k_2\neq k_3\nonumber\\
E_{\Sl(2)}^{(1,2)} = 3 n^2 \qquad 
E_{\Sl(2)}^{(1,3)}/E_{\Sl(2)}^{(1,2)} = -7/3 \qquad k_1=k_2=n, k_3=-2n\ ,
\ee
and we can easily verify that the agreement with table~\ref{NUM_SL(2)_1Loop}
is excellent.
\begin{table}[ht!]
\begin{eqnarray}
\begin{array}{|cccc|}
\hline
E_{\Sl(2)}^{(1,2)} & E_{\Sl(2)}^{(1,3)} & E_{\Sl(2)}^{(1,3)}/E_{\Sl(2)}^{(1,2)}  
                                  & (k_1,k_2,k_3) \\
\hline
1+1.2\times 10^{-9}	&-2-3.1\times 10^{-7}	&-2-3.1\times 10^{-7}	& (1,0,-1)	 \\
3-7.6\times 10^{-9}	&-7+1.9\times 10^{-6}   &-7/3+6.3\times 10^{-7} & (1,1,-2)	 \\	
3-7.6\times 10^{-9}	&-7+1.9\times 10^{-6}	&-7/3+6.3\times 10^{-7}	&  (-1,-1,2)	 \\
4-2.8\times 10^{-7}	&-8+6.9\times 10^{-6}   &-2+1.7\times 10^{-6}   &  (2,0,-2)	 \\
7-2.9\times 10^{-7}	&-14+7.1\times 10^{-5}	&-2+1.0\times 10^{-5}	&  (1,2,-3) 	 \\
7-2.9\times 10^{-7}	&-14+7.1\times 10^{-5}	&-2+1.0\times 10^{-5}	&  (-1,-2,3)	 \\
9-4.1\times 10^{-7}	&-18+1.0\times 10^{-4}	&-2+1.0\times 10^{-5}	&  (3,0,-3)	 \\
12+8.4\times 10^{-7}	&-28-1.5\times 10^{-4}	&-7/3-1.2\times 10^{-5}	&  (2,2,-4)	 \\
12+8.4\times 10^{-7} 	&-28-1.5\times 10^{-4}	&-7/3-1.2\times 10^{-5}	&  (-2,-2,4)	 \\
13-7.0\times 10^{-6}	&-26+1.7\times 10^{-3}	&-2+1.3\times 10^{-4}	&  (1,3,-4) 	 \\
13-7.0\times 10^{-6}	&-26+1.7\times 10^{-3}	&-2+1.3\times 10^{-4}	&  (-1,-3,4)	 \\
16-1.4\times 10^{-6}	&-32+3.9\times 10^{-4}	&-2+2.4\times 10^{-5}	&  (4,0,-4)	 \\
19-7.5\times 10^{-6}	&-38+2.2\times 10^{-3}	&-2+1.1\times 10^{-4}	&  (2,3,-5)	 \\
19-7.5\times 10^{-6}	&-38+2.2\times 10^{-3}	&-2+1.1\times 10^{-4}	&  (-2,-3,5)	 \\
21-3.4\times 10^{-6}	&-42+8.8\times 10^{-4}	&-2+4.2\times 10^{-5}	&  (1,4,-5)	 \\
21-3.4\times 10^{-6}	&-42+8.8\times 10^{-4}	&-2+4.2\times 10^{-5}	&  (-1,-4,5)	 \\
\hline
\end{array} \nonumber
\end{eqnarray}
\caption{Scaling limit of numerical spectrum of three-impurity
	$\Sl(2)$ sector at one-loop in $\lambda$}
\label{NUM_SL(2)_1Loop}
\end{table}

The extrapolated predictions can again be checked against those of 
the corresponding one-loop Bethe ansatz equations.
In the $\Sl(2)$ sector the highest weight is $-1/2$: the
Dynkin diagram therefore has coefficient $V_{\Sl(2)} = -1$ and the Cartan
matrix is $M_{\Sl(2)}=2$.  The Bethe equations (\ref{BAEFull},\ref{BAE2}) 
thus reduce to
\be
&&\kern-25pt
	\left(\frac{u_{i}-i/2}{u_{i}+i/2}\right)^L
	= \prod_{j\neq i}^{n}\left(
	\frac{u_{i} - u_{j} + i}{u_{i}-u_{j} - i}
	\right)
\label{BAESL2_1}
\\
&&
\kern+10pt	1 = \prod_{i}^{n}\left(
	\frac{u_{i} - {i}/{2}}{u_{i}+ {i}/{2}}
	\right)\ .
\label{BAESL2_2}
\ee
Apart from a crucial minus sign, this is identical to the $\su(2)$ Bethe equation (\ref{BAESU2_2}).
In the absence of bound states, eqn.~(\ref{BAESL2_1}) is satisfied by the following Bethe
roots:
\be
u_1 & = & -\frac{ 2(1+L)k_1^2 - (4+L)k_1 k_2 - (4+L)k_2^2}{2 \pi k_1(k_2^2+k_1 k_2-2k_1^2)}+O(L^{-1})
\nn\\
u_2 & = & -\frac{ 2(1+L)k_2^2 - (4+L)k_1 k_2 - (4+L)k_1^2}{2 \pi k_2(k_1^2+k_1 k_2-2k_2^2)}+O(L^{-1})
\nn\\
u_3 & = & -\frac{2(1+L)k_1^2 + (8+5L)k_1 k_2 + 2(1+L)k_2^2}{2\pi (k_1+k_2)(2k_1+k_2)(k_1+2k_2)}
	+O(L^{-1})\ .
\ee
Using eqn.~(\ref{betheenergy1L}), we obtain
\be
E_{\Sl(2)}^{(2)}(k_1,k_2) = \frac{\lambda}{L^3}\left(k_1^2 + k_1 k_2 + k_2^2\right)\left(L-2\right)
	+ O(L^{-4})
\qquad (k_1 \neq k_2 \neq k_3)\ .
\ee
For the bound state characterized by $k_1=k_2=n$ and $k_3=-2n$, 
the Bethe roots are
\be
u_1  &=&  \frac{7-3\sqrt{L}+3L}{6\pi n}+O(L^{-1/2}) \nn\\
u_2 &=& \frac{7+3\sqrt{L}+3L}{6\pi n}+O(L^{-1/2})  \nn\\
u_3 &=& -\frac{4+3L}{12\pi n}+O(L^{-1/2})\ ,
\ee
with spin chain energy
\be
E_{\Sl(2)}^{(2)}(n) = \frac{\lambda n^2}{L^3}(3L-7)
	+ O(L^{-4})  \qquad (k_1 = k_2 = n,\ k_3=-2n)\ .
\ee
These results again agree with the numerical results in table~\ref{NUM_SL(2)_1Loop}
(and match the corresponding three-impurity string theory predictions in
\cite{Callan:2004ev}).

\section{Conclusions}
We have demonstrated that the virial expansion of the ${\cal N}=4$ SYM 
spin chain Hamiltonian for small impurity number provides a simple and reliable
method for computing exact anomalous dimensions of multi-impurity operators 
at small scalar $R$-charge (chain length) and estimating with great precision the 
near-BMN scaling behavior of these dimensions as the $R$-charge becomes large.  
The latter application, which is suited to direct comparison 
of gauge theory predictions with corresponding results on the string side of
the AdS/CFT correspondence, works well for three-impurity operators 
to three-loop order in $\lambda$  in the $\su(2)$ sector 
(the order to which the $\su(2)$ Hamiltonian is known definitively).  
Specifically, the numerical predictions from the virial approach 
for the near-BMN scaling coefficients 
($E_{\su(2)}^{(1,2)},\ E_{\su(2)}^{(1,3)},\ 
E_{\su(2)}^{(2,4)},\ E_{\su(2)}^{(2,5)},\ 
E_{\su(2)}^{(3,6)}$ and $E_{\su(2)}^{(3,7)}$) 
match corresponding results from the $\su(2)$ long-range Bethe ansatz to three-loop order,
and agree with near-plane-wave 
string theory predictions to two loops (the disagreement with string theory
at three loops is by now an expected outcome in these studies).
We also find convincing agreement near the BMN limit between the virial approach
and the Bethe ansatz results at one-loop order in the closed $\Sl(2)$ and $\su(1|1)$
subsectors.  As a side result we have found 
in the $\su(1|1)$ sector an {\it exact} (in chain length) agreement between the Bethe ansatz 
and the virial expansion for one-loop operator dimensions with arbitrary 
impurity number (this was only possible because the Bethe equations can be
solved exactly in this subsector for any number of impurities).
There are currently no higher-loop Bethe ans\"atze for the $\Sl(2)$ and $\su(1|1)$
systems, however, so in this sense our numerical predictions go beyond 
the current state of Bethe ansatz technology (see \cite{Minahan:2004ds} for further 
developments of higher-loop gauge theory physics in non-$\su(2)$ sectors).  
It would be very interesting to find a general long-range Bethe equation 
appropriate for ${\cal N}=4$ SYM at higher loop-order in $\lambda$, 
both for comparison with string predictions and with the virial approach 
studied here.  

\section*{Acknowledgements}
This work was supported in part by US Department of Energy 
grants DE-FG02-91ER40671 (Princeton) and DE-FG03-92-ER40701 (Caltech).



\end{document}